\documentclass[]{aa}  
\makeatletter
\renewcommand*\aa@pageof{, page \thepage{} of \pageref*{LastPage}}
\makeatother
\usepackage{graphicx}
\usepackage{ulem}
\usepackage{hyperref}

\usepackage{txfonts}
\usepackage{todonotes}
\usepackage{caption}
\usepackage{amssymb}

\usepackage{dcolumn}
\usepackage{bm} 

\usepackage{float}

\usepackage{titlesec}
\newcommand{\kidsdata}{\texttt{KiDS-1000}}

\hypersetup{colorlinks=true, linkcolor=blue, citecolor = blue, urlcolor=cyan}

\usepackage{soul}

\let\oldpageref\pageref
\renewcommand{\pageref}{\oldpageref*}
\begin{document}

\title{Constraining hot dark matter sub-species with weak lensing and the cosmic microwave background radiation}
\titlerunning{Constraining Hot Dark Matter Sub-Species with WL and the CMB}
  \author{Fabian Hervas Peters\thanks{fabian.hervaspeters@cea.fr}\inst{1,2}
          \and 
          Aurel Schneider\inst{1}
          \and
          Jozef Bucko\inst{1}
          \and
          Sambit K. Giri\inst{1,3}
          \and
          Gabriele Parimbelli\inst{4,5,6}
          }

  \institute{Institute for Computational Science, University of Zurich, Winterthurerstrasse 190, 8057 Zurich, Switzerland \and
  Université Paris-Saclay, Université Paris Cité, CEA, CNRS, Astrophysique, Instrumentation et Modélisation Paris-Saclay, 91191 Gif-sur-Yvette, France \and 
  Nordita, KTH Royal Institute of Technology and Stockholm University, Hannes Alf\'vens v\"ag 12, SE-106 91 Stockholm, Sweden \and 
  Institute of Space Sciences (ICE, CSIC), Campus UAB, Carrer de Can Magrans, s/n, 08193 Barcelona, Spain \and 
  Dipartimento di Fisica, Universit\`a degli studi di Genova, and INFN-Sezione di Genova, via Dodecaneso 33, 16146 Genova, Italy \and
  SISSA, International School for Advanced Studies, Via Bonomea 265, 34136 Trieste, Italy
    }
    
  \date{Received: XX, XX, XXXX. Accepted: YY, YY, YYYY, Report Number: NORDITA 2023-032}


 
  \abstract
   {
    Although it is well known that the bulk of dark matter (DM) has to be cold, the existence of an additional sub-dominant, hot species remains a valid possibility. In this paper we investigate the potential of the cosmic shear power spectrum to constrain such a mixed (hot plus cold) DM scenario with two additional free parameters, the hot-to-total DM fraction ($f_{\rm hdm}$) and the thermal mass of the hot component ($m_{\rm hdm}$). Running a Bayesian inference analysis for both the Kilo-Degree Survey cosmic shear data (\kidsdata) as well as the cosmic microwave background (CMB) temperature and polarisation data from {\tt Planck}, we derive new constraints for the mixed DM scenario. We find a 95 per cent confidence limit of $f_{\rm hdm}<0.08$ for a very hot species of $m_{\rm hdm}\leq20$ eV. This constraint is weakened to $f_{\rm hdm}<0.25$ for $m_{\rm hdm}\leq80$ eV. Scenarios with masses above $m_{\rm hdm}\sim200$ eV remain unconstrained by the data. Next to providing limits, we investigate the potential of mixed DM to address the clustering (or $S_8$) tension between lensing and the CMB. We find a reduction of the 2D ($\Omega_m - S_8$) tension from 2.9$\sigma$ to 1.6$\sigma$ when going from a pure cold DM to a mixed DM scenario. When computing the 1D Gaussian tension on $S_8$ the improvement is milder, from 2.4$\sigma$ to 2.0$\sigma$.
    
   } 


   \maketitle
%

\section{Introduction}
\label{sec:intro}
During the early 1980s, neutrino particles were considered as dark matter (DM) candidates due to the mounting evidence of non-zero neutrino masses from particle physics experiments \citep{zeldovich_giant_1982, bond_how_1983}. This prompted the development of a top-bottom approach to structure formation, where massive "Zel'dovich Pancakes" were theorised to collapse or fragment into halos and galaxies. Such an approach was thought to allow for galaxy formation despite the substantial free-streaming properties of neutrino particles which counteract the gravitational clustering. However, the neutrino hot dark matter (HDM) model was quickly abandoned in favour of the cold dark matter (CDM) paradigm which predicted a more gradual build-up of galaxies in much better agreement with observations \citep{blumenthal_formation_1984}.

While CDM became a main component of the widely accepted $\Lambda$CDM theory, small-scale inconsistencies left room for hot and warm sub-components of DM. These inconsistencies included the missing satellite problem, relating to the mismatch between predicted DM sub-haloes and observed satellite galaxies, and the cusp-core problem, which involved the discrepancy between cuspy halo profiles from gravity-only simulations and cored profiles of observed dwarf galaxies \citep{moore_dark_1999, de_blok_core-cusp_2010}. One popular hypothesis to address these issues was the existence of a warm or hot DM sub-components \citep[e.g.][]{boyanovsky_dark_2008,Anderhalden:2012jc}. However, subsequent observations have shown this solution to be in tension with other data from the Lyman-$\alpha$ forest \citep{viel_warm_2013,markovic_lyman-_2014} and Milky Way satellite counts \citep{schneider_warm_2014}.

Today it is well established that the missing satellite and cusp-core problems can be alleviated by baryonic feedback effects and do not require additional modifications of the DM sector \citep[e.g.][]{brooks_baryonic_2013,del_popolo_review_2021}. However, in recent years another clustering tension has appeared between the $S_8$ clustering parameter from the CMB experiment {\tt Planck} and the stage-III lensing surveys {\tt KiDS} \cite[Kilo Degree Survey,][]{heymans_kids-1000_2021}, DES \cite[Dark Energy Survey,][]{amon_consistent_2022,des_collaboration_dark_2022}, or HSC \citep[Hyper Suprime Camera,][]{hsc_collaboration}. The $S_8$ parameter is defined as
\begin{equation}
S_8=\sigma_8\sqrt{\Omega_m/0.3},
\end{equation}
which includes the cosmological parameters $\Omega_m$ and $\sigma_8$, describing the matter abundance and the clustering amplitude, respectively. 

Attempts to alleviate the $S_8$ tension are often based on extensions to the $\Lambda$CDM framework that lead to modifications of the clustering process between the last scattering surface of the CMB and today. A few examples are decaying DM models \citep{fus_decaying_2022,abellan_linear_2021,bucko_constraining_2023,bucko_probing_2023}, new couplings between dark energy and DM \citep{poulin_sigma-8_2022}, or DM-baryon scattering \citep{he_s_8_2023}. 

In \cite{das_non-thermal_2022}, the effect of a hot sub-component of DM on the $S_8$ tension is studied at the linear level, assuming observations from the CMB together with low-redshift galaxy clustering probes. In this paper, we considered the same physical scenario, but we modeled the full non-linear process of structure formation. This allowed us to predict at the same time the CMB and the weak-lensing signal, obtaining a self-consistent test of the mixed DM scenario from these observations.

Another important goal of the present work was to use data from {\tt KiDS} and {\tt Planck} to derive new constraints on the particle mass ($m_{\rm hdm}$) and the fraction of hot-to-total DM
\begin{equation} \label{f_wdm}
f_{\text{hdm}}=\frac{\Omega_{\text{hdm}}}{\Omega_{\text{hdm}}+\Omega_{\text{cdm}}},
\end{equation}
where $\Omega_{\rm cdm}$ and $\Omega_{\rm hdm}$ are the abundances of the cold and hot components, respectively. In particular, we are interested in models with a HDM particle mass in the eV to keV range. In this mass regime, not many constraints currently exist as previous investigations have focused on either the sub-eV \citep{planck_coll_planck_2020} or the keV mass scales \citep{boyarsky_lyman-alpha_2009,Schneider:2014rda, baur_constraints_2017}.

In general, an additional HDM component leads to a suppression of the matter power spectrum at small scales \citep{viel_warm_2013}. This suppression is caused by the free-streaming of the hot component and depends on both the particle mass and the momentum distribution \citep[e.g.][]{boyarsky_lyman-alpha_2009,Merle:2015vzu}. Depending on the hot-to-total DM fraction ($f_{\rm hdm}$), the suppression can either have 
a steep cutoff (if $f_{\rm hdm}$ is close to one) or it can be more gradual and shallow (if $f_{\rm hdm}$ remains close to zero).

The MDM terminology is convenient in cosmological studies as it allows us to group together a wide variety of differently motivated theoretical models into a single phenomenological framework. Some prospective candidates in this mass range are the gravitino \citep[e.g.]{osato_cosmological_2016} 
and the sterile neutrino \citep{dodelson_sterile_1994, shi_new_1999}. The latter has been studied extensively for a variety of different production mechanisms, such as non-resonant mixing, resonant mixing or production via early decays \citep[see][for a review]{Drewes:2016upu}.
Since these particles are never in thermal equilibrium, their momenta cannot be described by a Maxwell-Boltzmann distribution. As a consequence, they exhibit some differences in the way they suppress the matter power spectrum. However, in many cases, a simple re-mapping of the particle mass allows us to interpret the suppression in terms of the standard thermal particles at a sufficient accuracy \citep{abazajian2006production,Merle:2014xpa,bozek_resonant_2016}.

The same occurs in the ultra-light axion scenario with one or several axion-like particles \citep{Marsh:2013ywa,hlozek_search_2015,giri_imprints_2022,vogt_improved_2023,rogers_ultra-light_2023}. In principle, these bosonic DM particles exhibit very different dynamics at the scale of their particle wavelengths. However, for many applications, the results can be brought into reasonable agreement with the fermionic mixed DM scenario  \citep{Hui:2016ltb}. 

Finally, many interacting DM scenarios have a similar effect on structure formation as the simple MDM model. This is especially true for the case where potential interactions of the dark sector are restricted to the early universe as is the case for most of the ETHOS \citep{cyr-racine_ethoseffective_2016} parameter space, for example. In this particular, simplest non-minimal scenario, the DM particle is allowed to interact with a mediator particle playing the role of dark radiation. The early-universe interactions between the DM and dark radiation particles cause a suppression of the power spectrum at small scales including, in some cases, dark acoustic oscillation features. As these features quickly disappear at nonlinear scales \citep{Schaeffer:2021qwm}, the ETHOS interaction framework often resembles the mixed DM case \citep{Archidiacono:2019wdp}.

In summary, the analysis presented in this paper is, strictly speaking, valid for the case of a mixed DM scenario where the hot particles undergo free streaming caused by a momentum distribution that is identical or close to a Maxwell-Boltzmann distribution. The prime example of such a model is CDM combined with a sterile neutrino produced in the standard way via the Dodelson-Widrow scenario. However, many other DM models mentioned above will yield very similar results and some of the conclusions can be transferred to other scenarios.


In the next section, we will describe our data analysis modeling method. In Section~\ref{sec:results}, we will present our findings, and finally, we will conclude our study in Section~\ref{sec:conclusion}.

\section{Methods}
In this section, we summarise our pipeline to predict the cosmic shear and the CMB data for both the CDM and MDM models. Note that the description remains at a general level, more details can be found in \cite{schneider_constraining_2022} and \cite{bucko_constraining_2023}.

\begin{figure*}
\centering

\includegraphics[width=0.85\textwidth]{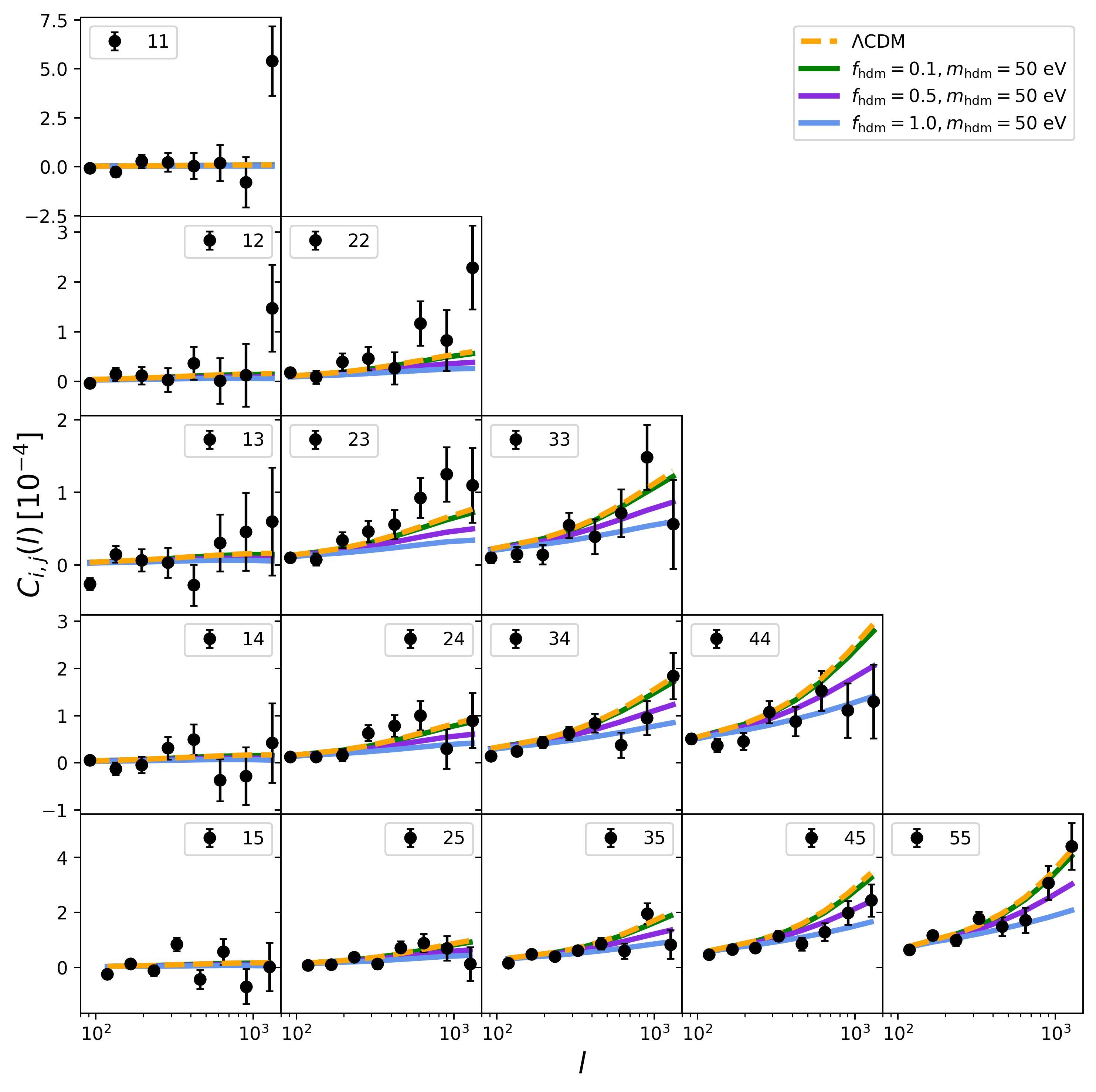}
\caption{\label{fig:mwdm_cl} Auto and cross angular band power spectra of the \kidsdata{} data set separated in five tomographic redshift bins (black data points). The data is obtained from \citet{asgari_kids-1000_2021}. The lines correspond to predictions assuming a $\Lambda$CDM model (orange dashed) and three $\Lambda$MDM models with fixed thermal mass ($m_{\rm hdm}=50$  keV) and increasing hot-to-total DM fraction ($f_{\rm hdm}=$0.1, 0.5, 1.0) in green, purple, and blue. All plotted models are run assuming the Planck-18 best-fitting cosmology. }
\end{figure*}

\subsection{Cosmic Shear Signal}\label{sec:CSsig}
For this analysis, we relied on the band power data from the \kidsdata{}  cosmic shear observations published 
in \citet{asgari_kids-1000_2021}. A detailed description of how to obtain predictions for the band power data is given in \citet{joachimi_kids-1000_2021}. See also Sec.~3.1. of \citet{schneider_constraining_2022} for more details. In general, the band power data is obtained via an integral over the angular power spectrum multiplied by the band filter functions \citep{joachimi_kids-1000_2021}. The angular power spectrum is obtained via the Limber approximation \citep{limber_analysis_1953}
\begin{equation}
C_{i,j}^{A,B}(\ell)=\int_0^{\chi_H}\frac{W_i^A(\chi)W_j^B(\chi)}{f_\kappa(\chi)^2}P_{\rm tot}\left(\frac{\ell}{\chi},\chi\right) \ d\chi
\end{equation}
where the $i,j$ subscripts refer to the tomographic redshift bins and $P_\mathrm{tot}$ is the total matter power spectrum.

The window functions $W_A$ and $W_B$ describe the lensing (G) and intrinsic alignment (I) weights, [A,B] $\in$ [I,G]. In a flat universe, they are given by
\begin{eqnarray}
W^G_{i}(\chi) &=& \frac{3\Omega_m H_0^2}{2c^2}  \chi(1+z) \int_{\chi}^{\chi_H} d\chi' n_{i,S}(\chi') \frac{(\chi'-\chi)}{\chi'} \ ,\nonumber\\
W_i^I(\chi)&=-&A_{IA} \frac{C_1 \rho_{\text{cr}}\Omega_m}{D(z)}n_{i,S}(\chi) \ ,
\end{eqnarray} 
where $\chi$ is the comoving distance, $H_0$ the Hubble parameter, $\rho_{cr}$ the critical density, and $D(z)$ the growth factor. The redshift distribution of galaxies, $n_{i,S}(z)$, is obtained from \citet{asgari_kids-1000_2021} assuming 5 tomographic bins. The parameter $A_{IA}$ describes the amplitude of the intrinsic alignment effect -- assuming the Non-Linear Alignment (NLA) Model described in \citet{hirata_intrinsic_2010} and \citet{hildebrandt_kids-450_2017} -- and was kept as a free parameter. The  $C_1$ value was calibrated to match SuperCOSMOS Sky Surveys (SSS) observations \citep{brown_measurement_2002}, such that the free parameter $A_{IA}$ has value of unity. In accordance with \cite{heymans_kids-1000_2021}, any redshift dependence of the intrinsic alignment effect was ignored for this analysis.

The total nonlinear matter power spectrum is given by
\begin{equation}\label{PStot}
P_{\rm tot} (k,z) =  S_{\rm baryon}(k,z)\times S_{\rm mdm}(k,z)\times P_{\rm NL}(k,z) \ ,
\end{equation}
where $P_{\rm NL}(k,z)$ is the nonlinear, gravity-only, matter power spectrum obtained with the {\tt revised\_halofit} method of \citet{takahashi_revising_2012}. Note that $P_{\rm NL}(k,z)$ only depends on the cosmological, but not on any astrophysical or DM parameters. The function $S_{\rm baryon}(k,z)$ refers to the suppression due to the baryonic feedback effect. We modeled this effect using the emulator {\tt BCemu} \citep{giri_emulation_2021}, which is based on the baryonification method \citep{schneider_new_2015,Schneider:2018pfw}. 
This emulator depends on the cosmic baryon fraction ($f_b=\Omega_b/\Omega_m$) plus seven free baryonic parameters. However, instead of varying all of them \citep[as in][]{schneider_constraining_2022}, we only vary the three baryonic parameters $\lbrace\log_{10}M_c, \eta_d, \theta_j\rbrace$ along with $f_b$. This reduced setup of parameters has been shown in \citet{giri_emulation_2021} to provide accurate fits to all investigated hydrodynamical simulations. We show posteriors of baryonic feedback parameters in Appendix~\ref{appendix:baryons}

The function $S_{\rm mdm}(k,z)$ characterizes power spectrum changes due to hot/warm DM using an improved emulator derived from \citet{parimbelli_mixed_2021}. This new version is trained on a larger set of simulations (100 vs. 74) across a broader parameter range $(f_\mathrm{\rm hdm}, m_\mathrm{\rm hdm})$, encompassing masses as low as 0.03 keV. Initial conditions are established through the fixed-and-paired technique \citep{angulo2016cosmological}, with simulations conducted within a 120 Mpc/$h$ box. This ensures precision convergence of better than 1\% for $k\lesssim 10 \ h/$Mpc, effectively maintaining a link with the linear regime.

The emulator obviates the need for computing a separate linear MDM spectrum, as it yields the non-linear MDM power spectrum by adjusting a non-linear $\Lambda$CDM spectrum. In the parameter space accepted by the \kidsdata \ data, $\sigma_8$ discrepancies between $\Lambda$CDM and MDM calculations are negligible. Furthermore, \citet{parimbelli_mixed_2021} demonstrated that the suppression induced by MDM remains cosmology-independent within a 2\% range across various $\Omega_\mathrm m$ and $\sigma_8$ values.

Note that in our modelling we implicitly assumed that the baryonic feedback and MDM suppression effects are independent from each other. As a consequence, the functions $S_{\rm baryon}(k,z)$ and $S_{\rm mdm}(k,z)$ can be multiplied as shown in Eq.~(\ref{PStot}) greatly simplifying the analysis. The validity of this approximation has been confirmed in \citet{parimbelli_mixed_2021} with the help of baryonified CDM and MDM simulations. It turns out that multiplying the two suppression functions only adds a small, sub-percent error to the full power spectrum. This is significantly smaller than the estimated error of power spectrum estimator itself.

In Fig.~\ref{fig:mwdm_cl} we show the band power spectrum from the \kidsdata{} analysis together with predictions for MDM models with fixed particle mass ($m_{\rm hdm}=50$ eV) and a varying fraction ($f_{\rm hdm}=0.1,\,0.5,\,1$) using the pipeline summarised above. All other cosmological parameters are kept at the Planck 18 values. The plot shows that the higher the ratio of hot-to-cold DM, the more the band power is suppressed. Furthermore, the suppression is more pronounced at higher $l$-modes, which correspond to smaller physical scales, as well as at higher redshift bin numbers indicating a stronger suppression at larger redshift. Note that although the effect of mixed DM is very well visible, potential degeneracies with cosmology and baryonic physics may strongly reduce the effect.

The angular power spectra were calculated using the {\tt PyCosmo} package \citep{refregier_pycosmo_2017,tarsitano_predicting_2020}. The prior ranges are summarised in Table~\ref{tab:priors}. More details about the selection of the priors are provided in \cite{bucko_constraining_2023}.
Note that for comparison, we also ran an analysis assuming the standard $\Lambda$CDM model. We thereby find good agreement with the \kidsdata{} results from \citet{asgari_kids-1000_2021} and \citet{schneider_constraining_2022} validating our pipeline.

\subsection{CMB data}
We extended our exploration beyond cosmic shear and delved into the impact of MDM on the CMB temperature and polarization data. For this analysis, we employed the {\tt Planck-lite-py} module from \citet{prince_data_2019}, using the Planck-18 TTTEEE likelihood provided in the package, including low-$\ell$ bins. The Planck-18 likelihood was coupled with the Boltzmann solver {\tt CLASS} \citep{blas_cosmic_2011}. {\tt CLASS} features an integrated option for additional hot and warm DM sub-species \citep{lesgourgues_cosmic_2011}.  While this study was conducted with the thermal relics mass $m_{\rm{therm,wdm}}$, a conversion is possible to sterile neutrino mass $m_{\nu_s}$ through following expression \citep[provided in][]{bozek_resonant_2016}: 
\begin{equation} \label{m_nu_conv}
    m_{\nu_,s}=3.90 \ \rm{keV} \ \left(\frac{m_{\rm{thermal}}}{1keV}\right)^{1.294}\left(\frac{f_{\rm{wdm}}\Omega_{\rm{DM}}h^2}{0.1225}\right)^{-1/3} \ .
\end{equation}
The {\tt CLASS} Boltzmann solver takes the sterile neutrino mass as an input for the {\tt m\_ncdm} parameter

\begin{figure*}[htp]
\includegraphics[height=6.5cm,trim=0.cm 0.1cm 0.8cm 0.5cm, clip]{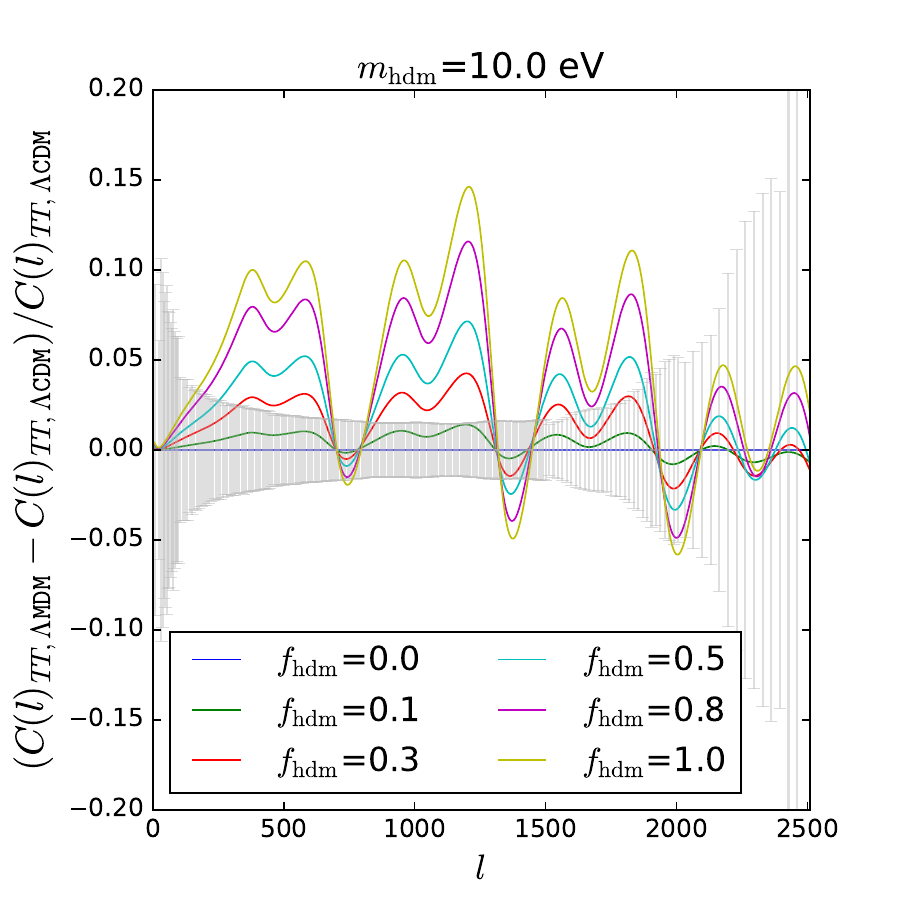}
\includegraphics[height=6.5cm,trim=1.7cm 0.1cm 0.8cm 0.5cm, clip]{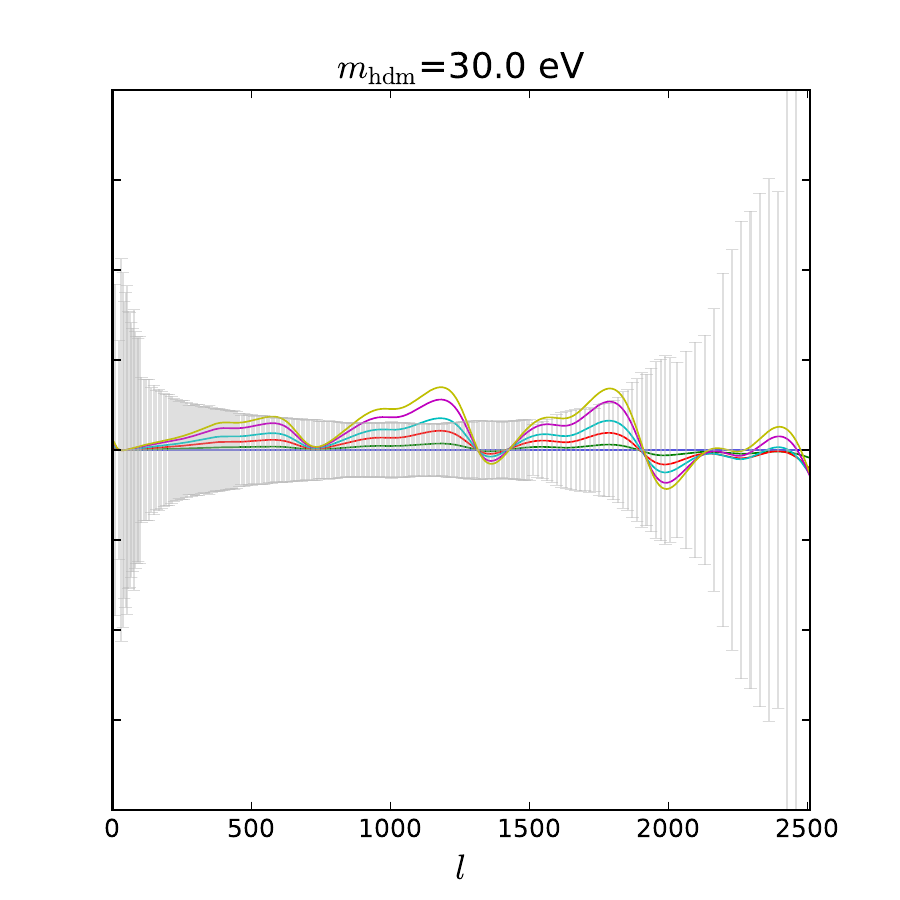}
\includegraphics[height=6.5cm,trim=1.7cm 0.1cm 0.8cm 0.5cm, clip]{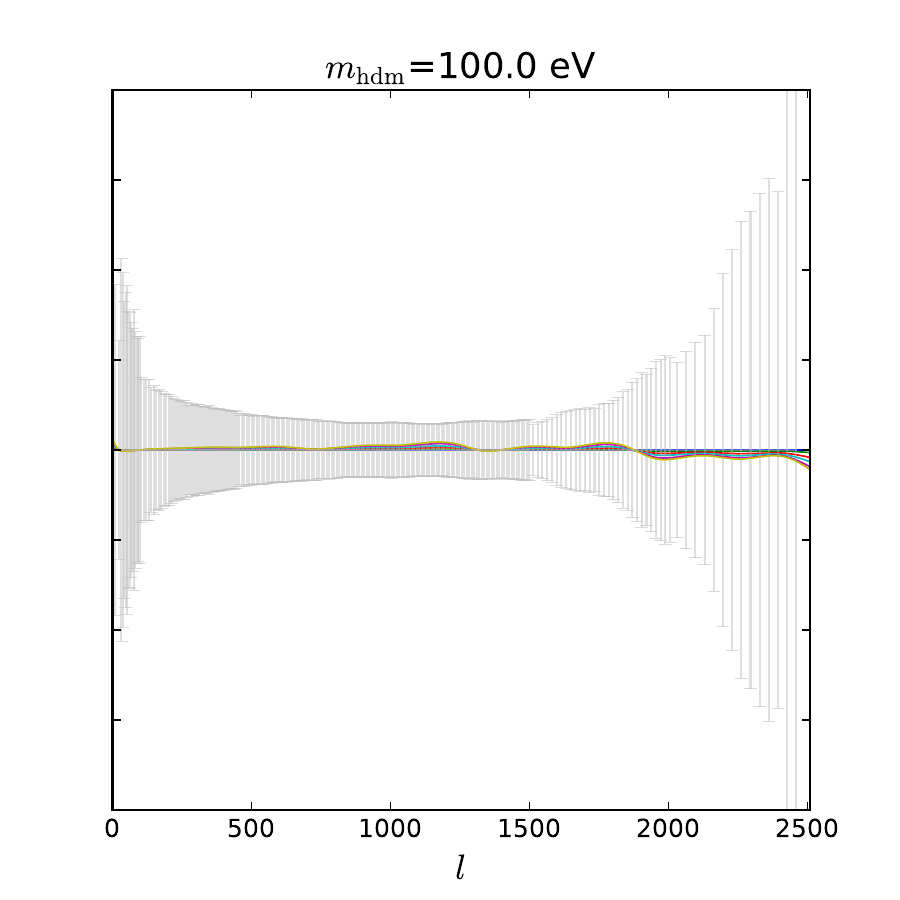}
\caption{Residual angular power spectra     of the CMB temperature fluctuations from Planck-18 assuming a mixed DM model with varying hot-to-total DM fractions (coloured lines) and a thermal particle mass of $m_{\rm hdm}=10$ eV (left), $m_{\rm hdm}=30$ eV (middle), and $m_{\rm hdm}=100$ eV (right). Note that the blue line ($f_{\rm hdm}=0$) corresponds to the $\Lambda$CDM model. The grey error bars denote the 68\% confidence interval of the Planck-18 TT data.}
\label{fig:planck_cl}
\end{figure*}

In Fig.~\ref{fig:planck_cl}, we visualize the influence of an extra HDM sub-species across diverse particle masses and hot-to-total DM fractions. The three panels depict models for particle masses of $m_{\rm hdm}=\lbrace 10, 30,100\rbrace$ eV, demonstrating the diminishing effects on the $C(\ell)$ as the mass rises to $100$~eV. While the apparent increase in $C(l)_{\rm TT,\Lambda MDM}$ seems in conflict with the objective of lowering $S_8$, the larger peaks are due to a shift in the energy budget from matter to radiation when increasing $f_{hdm}$ for low masses. This decreases {$\omega_m/\omega_R$} therefore delaying the matter-radiation $z_{eq}$ which leaves a shorter time for the BAO peaks to decrease until recombination.

\begin{table}
\renewcommand{\arraystretch}{1.3}
\scalebox{0.85}{
\begin{tabular}{l|c|c}
  Parameters & Symbol& Prior \\
  \hline \hline 
  Cold DM energy density&$\omega_{\rm cdm}$ & [0.051, 0.255]\\
  Baryon energy density&$\omega_{\rm b}$ & [0.019, 0.026]\\
  Initial $P(k)$ amplitude&$\log(10^{10} A_{\rm s}) $ & [1.0, 5.0] \\
  Hubble constant &$h$&[0.5, 0.9]\\
  Spectral index &$n_s$ &a[0.8, 1.3]\\

  \hline 
  Optical depth &$\tau$&$\mathcal{N}$(0.0544, 0.007)\\
  \hline
  Intrinsic Alignment amplitude&$A_{IA}$&[-2, 2]\\
  First gas parameter ({\fontfamily{lmtt}\selectfont BCemu}) &$\log_{10}(M_c)$&[11.0, 15.0]\\
  Second gas parameter ({\fontfamily{lmtt}\selectfont BCemu})&$\theta_j$&[4.0, 6.0]\\
  Stellar parameter ({\fontfamily{lmtt}\selectfont BCemu}) &$\eta_\delta$&[0.05, 0.40]\\
  \hline
  HDM fraction &$f_{\text{hdm}}$&[0.0, 1.0]\\
  Thermal mass &$\log_{10}(1/m_{\text{hdm}}(\rm{keV}))$ &[-0.17, 2.00]
\end{tabular}}
\hspace{1.5cm}
\caption{Prior ranges used in the Planck-18 TTTEEE and \kidsdata \ analysis. We used flat priors for all the parameters excluding $\tau$, which has a Gaussian prior to recover the posteriors of the uncompressed analysis as explained in \citep{prince_data_2019}.
Horizontal lines distinguish: I) standard $\Lambda$CDM parameters, II) CMB only parameters, III) \kidsdata \ parameters, and IV) MDM extension parameters. }
\label{tab:priors}
\end{table}

\subsection{Bayesian inference analysis}

We performed a Bayesian inference analysis with a Monte-Carlo Markov Chain (MCMC) method implemented in the {\tt emcee} package \citep{foreman-mackey_emcee_2013}. 
This method determines the posterior probability $p(\pmb{\theta}|\pmb{d})$ on the interested parameters $\pmb{\theta}$, given the observation $\pmb{d}$.
To explore the vast multidimensional parameter space, our analysis was performed on the supercomputer facilities at the Swiss National Supercomputing Centre (CSCS), running in parallel on 128 CPU cores.
Our priors $p(\pmb{\theta})$ are mostly flat and wide as described in Table~\ref{tab:priors}.
Note that we have verified our analysis pipeline by reproducing the fiducial $\Lambda$CDM analysis from {\tt Planck}.

\section{Results}\label{sec:results}
Here we present our new constraints on the MDM model (Section~\ref{sec:constraints_mdm}) and its implications on the $S_8$ tension (Section~\ref{sec:impact_S8}).

\subsection{Constraints on MDM model}\label{sec:constraints_mdm}
\begin{figure}[!h]
\centering
\includegraphics[width=0.52\textwidth]{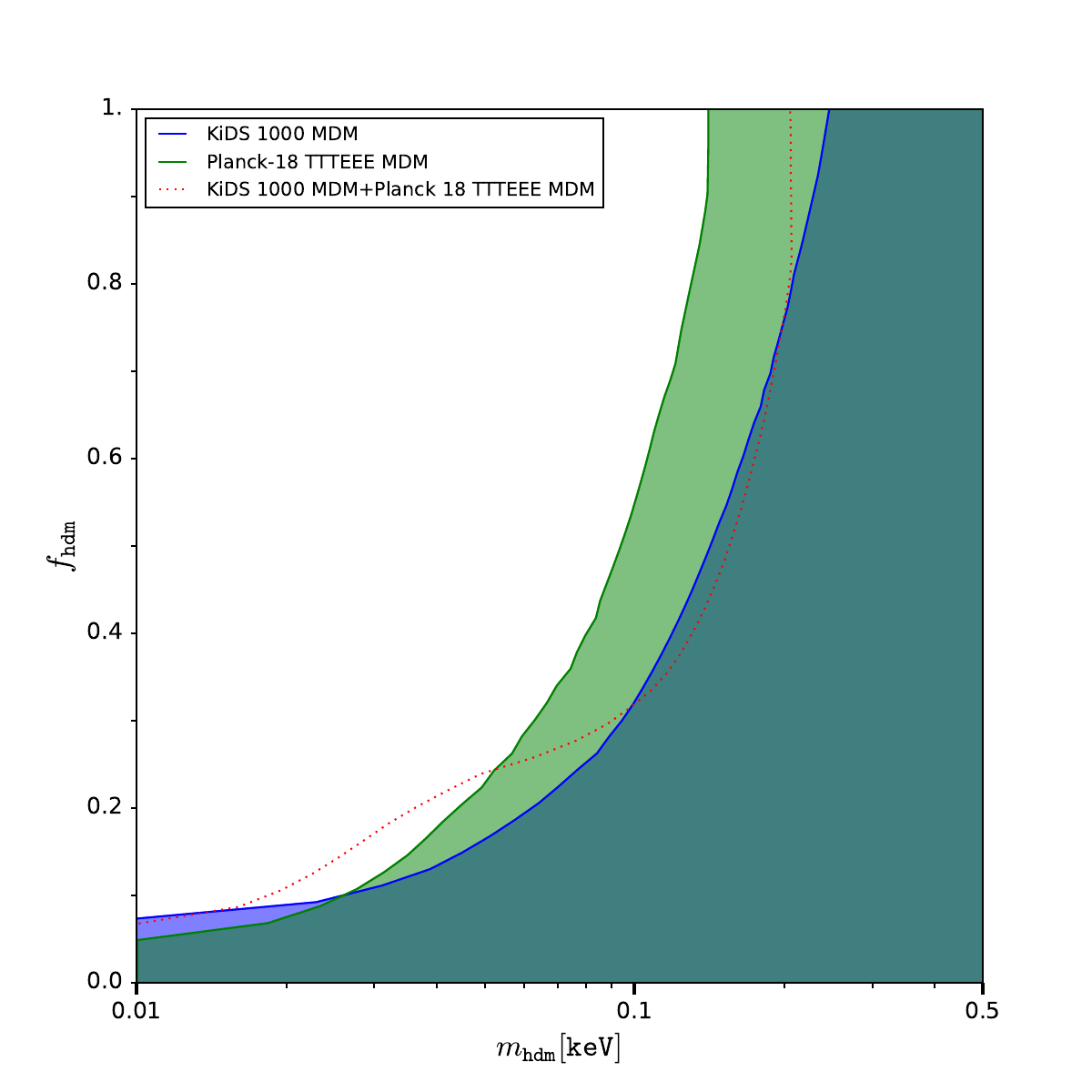}
\caption{Constraints on the MDM ($\Lambda$MDM) model from weak lensing (\kidsdata) in green, the CMB (Planck-18 TTTEEE) in blue, and the combined WL+CMB in red (dotted line). The$\Lambda$MDM model is parametrised by the thermal mass of the hot species ($m_{\rm hdm}$) and the fraction of hot-to-total DM ($f_{\rm hdm}$). The top-left corner of the plot corresponds to the excluded region. All contours are shown at the 95\% confidence level.}  
\label{fig:mass_limit} 
\end{figure}

The constraints on the MDM parameters were obtained from the full Bayesian inference chains by marginalising over all parameters except $m_{\rm hdm}$ and $f_{\rm hdm}$. We show the results of this analysis in Fig.~\ref{fig:mass_limit}. All constraints are provided at the 95 \% confidence level. For a pure $\Lambda$HDM scenario with $f_{\rm hdm}=1$, the cosmic shear and CMB analysis yield constraints of $m_{\rm th}<0.2$ keV and $m_{\rm th}<0.14$ keV, respectively. We should note that these limits are significantly weaker than other constraints from e.g. Milky-Way satellite counts or the Lyman-$\alpha$ forest.

Both the cosmic shear and CMB data are much more powerful in constraining models with small hot-to-total DM fractions. For a particle mass of $m_{\rm hdm}\leq 20$ eV we obtain limits of $f_{\rm hdm}<0.09$ and $<0.08$  from the WL and the CMB analysis. This means that a HDM particle can not make up more than 8-9\% of the total DM budget.

The constraints on $f_{\rm hdm}$ become weaker when going to larger particle masses. For the cosmic shear (and CMB) analysis we obtain the limits $f_{\rm hdm}<0.16$ ($<0.24$) for $m_{\rm hdm}\leq 50$ eV and $f_{\rm hdm}<0.3$ ($<0.55$) for $m_{\rm hdm}\leq 100$ eV. 

In Fig.~\ref{fig:mass_limit} we also show the constraints of the combined \kidsdata \ and Planck-18 analysis. They are slightly weaker than the limits obtained with the runs based on the individual weak-lensing cosmic shear and CMB data. This surprising result is caused by the existing clustering (or $S_8$) tension between the two data-sets. It turns out that the tension is compensated in the combined chain by allowing for slightly higher values of the hot-to-total DM fraction. 

Parts of the MDM parameter ranges investigated here were explored in the past using data from the Lyman-$\alpha$ forest \citep{boyarsky_lyman-alpha_2009, baur_constraints_2017} and Milky-Way satellite counts \citep{Schneider:2014rda}. These studies found constraints of $f_{\rm hdm}\sim 0.1-0.2$ for small particle masses which is only slightly weaker than our constraints. However, they only focused on the regime above $m_{\rm hdm}\sim 0.2$ keV and they did not perform a Bayesian inferences analysis including cosmological parameters. Furthermore, the Lyman-$\alpha$ studies assumed a IGM temperature-evolution that follows a power law which is known to yield very constraining results \citep{garzilli_how_2021}.

We have focused this work on the full exploitation of small scale information in large scale structure. While Lyman-$\alpha$ probe the structure formation at high redshift, CMB lensing probes the universe at a similar redshift range on larger scales. Recent data as published by ACT \citep{madhavacheril_atacama_2023} and SPT \citep{pan_measurement_2023} could therefore help to constrain low-mass particles with large free-streaming scales.

\subsection{Comparison to other work}
In a recent study by \cite{das_non-thermal_2022} a MDM model with additional neutrino-like sub-species was investigated at the linear level. The paper focused on Planck data combined with BAO measurements from the BOSS survey as well as the Pantheon SNIa catalogue data. An additional prior mimicking the $S_8$ measurement from \kidsdata{} \citep{heymans_kids-1000_2021} was included in the analysis. Note that their approach included results obtained under the premises of $\Lambda$CDM and can therefore only be used as an approximation.

Converting the model of \cite{das_non-thermal_2022} into our parametrisation, we find that their best-fit model (which was claimed to alleviate the $S_8$ tension) corresponds to $m_{\rm therm,hdm}=11.36$~eV and $f_{\text{hdm}}=0.08$ \citep[see table~III in][]{das_non-thermal_2022}. Interestingly, this point is right on the 95\% confidence level of our cosmic shear analysis (see Fig.~\ref{fig:mass_limit}), which means that our results neither confirm nor strongly disfavour the best fitting model found by \cite{das_non-thermal_2022}. For a description of the different ways to parameterize light relics, we refer the reader to \cite{acero_cosmological_2009}.

A recent forecast study by \citet{Schneider:2019snl, schneider_baryonic_2020} investigated the potential of the Euclid weak-lensing survey to constrain the MDM model. They found that Euclid will provide much stronger constraints on the mass and fraction of a potential HDM particle. Assuming a $\Lambda$CDM universe, the reported limits are $f_{\rm hdm}<0.01$ for $m_{\rm hdm}\lesssim 30$ eV (at the 95\% confidence level). This confirms that future, stage-IV lensing surveys will be able to detect hot particle sub-species even if they only make up a percent of the total DM budget.

Similarly to \cite{das_non-thermal_2022}, a recent study by \cite{rogers_5_2024} found preference for a MDM type suppression in the eBOSS Lyman-$\alpha$ forest data. The parameters they find to alleviate the tension of 4.9$\sigma$ to 1.34$\sigma$ is $\mathrm{log}(m_{\rm wdm})=1.01_{-0.44}^{+0.30}$ and $f_{\mathrm{wdm}}=0.0219_{-0.0042}^{+0.0030}$. The 10eV mass which was found to alleviate the tension is right at the edge of our prior, and the $f_{\mathrm{wdm}}$ fraction of 0.02 is comfortably inside our constraints. While the results presented in \cite{rogers_5_2024} seem coherent with what we have found, more stringent weak lensing data would be required to probe these low fractions of MDM.

\subsection{Impact on the $S_8$ tension} \label{sec:impact_S8}
A further goal of this work was to investigate the effects of the $\Lambda$MDM model on the $S_8$ tension. The gradual power suppression towards small scales caused by a subdominant HDM species could affect the nonlinear clustering by pushing the weak-lensing estimates for $S_8$ further up. This could then lead to a better agreement with the measurements from the CMB.

\begin{figure*}[!htb]
 \centering
    \includegraphics[width=6.5in,trim=0cm 1.5cm 0.cm 4cm]{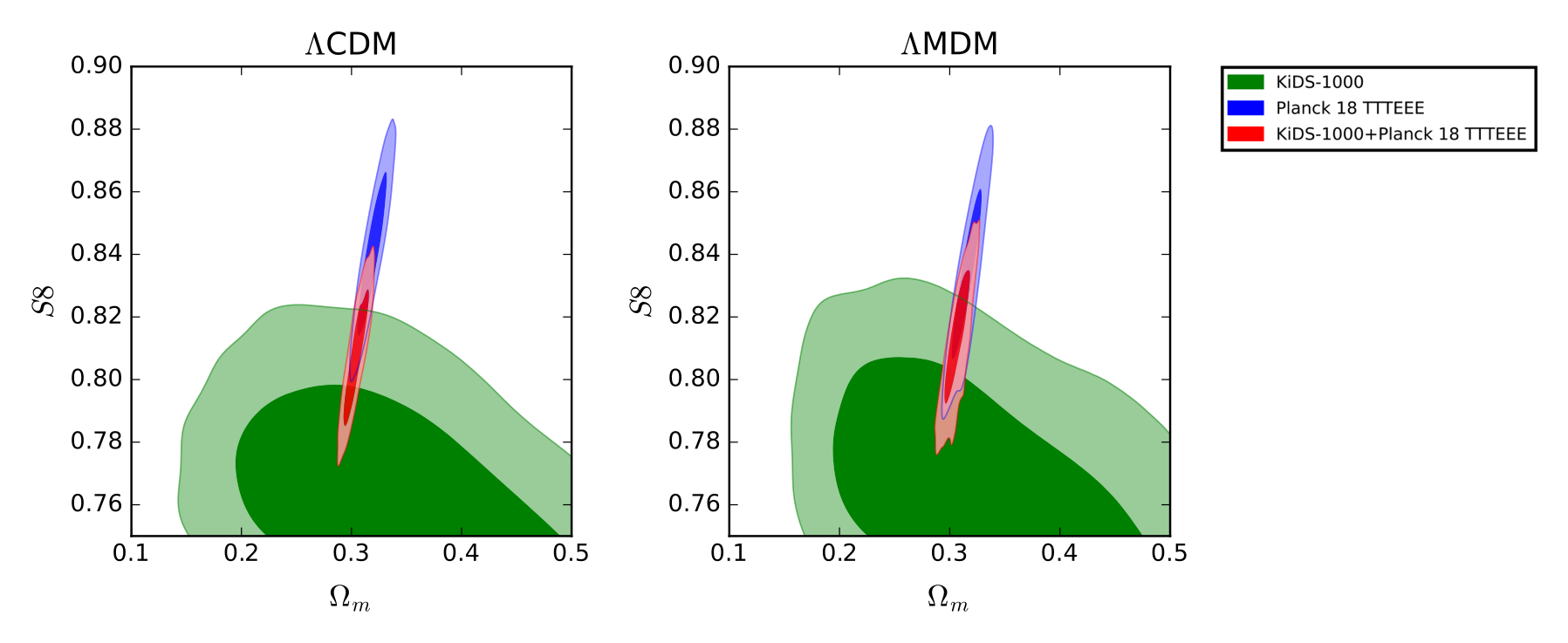} %
    
    \caption{Posterior contours in the $S_8-\Omega_m$ plane for the $\Lambda$CDM (left) and $\Lambda$MDM models (right) at 68 and 95 \% confidence level. The green, blue, and red  contours correspond to the cosmic shear (\kidsdata), the CMB (Planck-18 TTTEEE) and the combined analysis. 
}
\label{fig:S8}
\end{figure*}

In Fig.~\ref{fig:S8} we plot the two-dimensional posterior contours of $S_8$ and $\Omega_m$ for both the $\Lambda$CDM and $\Lambda$MDM scenarios. The weak-lensing posteriors (green) are indeed pushed to somewhat higher values of $S_8$ in the case of $\Lambda$MDM, but the shift is not very significant. At the same time the posterior from the CMB analysis (blue) moves slightly down, helping to alleviate the $S_8$ tension. For completeness in Fig.~\ref{fig:S8}, we also provide the results from the combined \kidsdata \ and Planck-18 TTTEEE analysis . The corresponding posteriors are shown in red. Not surprisingly, they lie between the results from the individual weak lensing and CMB analyses.
The posterior means for \kidsdata{}  are $S_8=0.749_{-0.029}^{+0.034}$ and $S_8=0.754_{-0.030}^{+0.034}$ for $\Lambda$CDM and $\Lambda$MDM respectively. For the Planck 18 data the values are $S_8=0.841\pm 0.017 $ and $S_8=0.832\pm 0.019 $ for $\Lambda$CDM and $\Lambda$MDM respectively.
We find that the remaining cosmological parameters are not modified by the change of model. 
We used two metrics to estimate the tension, which we refer to as 2D and 1D.

To calculate the 2D tension, we used the {\fontfamily{lmtt}\selectfont tensiometer} package \citep{raveri_non-gaussian_2021}. The package first computes the difference between the parameters in the chains and then estimates the probability of the shift using a Kernel Density Estimate algorithm (KDE) described in \cite{raveri_quantifying_2020}. Formally, we obtained a reduction of the tension from 2.9$\sigma$ in $\Lambda$CDM to 1.6$\sigma$ in the $\Lambda$MDM model (see also Table~\ref{tab:tension_metric}). 

When computing the gaussian 1D tension in $S_8$, we used the metric from \cite{asgari_kids-1000_2021} that is given as
\begin{eqnarray}
    \label{eq:S8-tension-def}
    & &\tau_{S_8} = \frac{S_8^{\rm CMB} - S_8^{\rm WL}}{\sqrt{\rm{Var}\left[S_8^{\rm CMB}\right] + \rm{Var}\left[S_8^{\rm WL}\right]}},
\end{eqnarray}
where $S_8^{\rm CMB}$ and $S_8^{\rm WL}$ are the most probable values derived from CMB and WL observations respectively. In the denominator, the variance of the same quantities are used. In this 1D framework, the tension is reduced from 2.4$\sigma$ to 2.0$\sigma$ only. 
While the tension values computed using the 1D and 2D estimates differ quite strongly, the non Gaussian shapes of the weak lensing contours tend to make us favour the 2D tension estimate based on parameter shifts.

\begin{table}
\centering
\renewcommand{\arraystretch}{1.3}
\scalebox{0.85}{
\begin{tabular}{l|c|c}
 & Planck-18 $\Lambda$CDM & Planck-18 $\Lambda$MDM \\
\hline \hline 
KiDS-1000 $\Lambda$CDM & 2.9 $\sigma$ & 1.7 $\sigma$ \\
KiDS-1000 $\Lambda$MDM & 1.7 $\sigma$ & 1.6 $\sigma$ \\
\end{tabular}}
\caption{2D tension metrics between Planck-18 and \kidsdata{} datasets for $\Lambda$CDM and $\Lambda$MDM models computed using parameter shifts in the $\Omega_m - S_8$ plane with the {\tt tensiometer} package.}
\label{tab:tension_metric}
\end{table}



\section{Conclusion} \label{sec:conclusion}
The presence of a sub-dominant HDM species, such as an additional neutrino particle, is a straight-forward extension of the $\Lambda$CDM scenario. In this paper, we explored the power of cosmic shear together with CMB temperature and polarisation data to constrain the mixed DM model ($\Lambda$MDM) which is parametrised by the (thermal) particle mass of the hot species ($m_{\rm hdm}$) and the hot-to-total DM fraction ($f_{\rm hdm}$). As observations we use the \kidsdata \ band power data from \citet{asgari_kids-1000_2021} as well as the Planck-18 TTTEEE data from \citet{planck_coll_planck_2020}.

We find new constraints on the $\Lambda$MDM parameters that are summarised in Fig.~\ref{fig:mass_limit}. At the 95\% confidence level, the hot-to-total DM fraction is limited to $f_{\rm hdm}<0.08$ for a hot species with $m_{\rm hdm}\leq 20$ eV. This limit is weakened to $f_{\rm hdm}<0.16$ for $m_{\rm hdm}\leq 50$ eV and $f_{\rm hdm}<0.30$ for $m_{\rm hdm}\leq 100$ eV. Scenarios with (thermal) particle masses beyond $m_{\rm hdm}\sim 200$ eV remain unconstrained by the weak lensing and CMB data.

Next to providing constraints, we also investigated the $S_8$ (or lensing) tension describing the clustering mismatch between cosmic shear probes and the CMB data. We find that the $S_8$ tension is decreased from 2.9$\sigma$ in $\Lambda$CDM to 1.6$\sigma$ in the $\Lambda$MDM model. This  improvement is caused by both a slight upper shift of the \kidsdata \ as well as a small downward shift of Planck-18 TTTEEE contours. The two dimensional $\Omega_m-S_8$ posterior contours are illustrated in Fig.~\ref{fig:S8}. Using 1D estimates the reduction in tension is milder, from 2.4$\sigma$ to 2.0$\sigma$. While these estimates are quite different we tend to favour the 2D parameter shift methods as it was designed to capture non-Gaussian features in the posteriors.

In the near future, stage-IV lensing surveys such as Euclid will allow us to further probe the $\Lambda$MDM model. In particular, it will be possible to constrain models where only about 1\% of the DM sector is made of a hot particle. This is about a factor of ten improvement with respect to current weak-lensing observations.

\begin{acknowledgements}

For the plotting routines we used the {\fontfamily{lmtt}\selectfont GetDist } \citep{lewis_getdist_2019} and {\fontfamily{lmtt}\selectfont matplotlib} \citep{hunter_matplotlib_2007} packages. 

We also thank Giovanni Aric\`o and Andrej Obuljen for informing discussions.
This work is supported by the Swiss National Science Foundation under grant number PCEFP2$\_$181157.
Nordita is supported in part by NordForsk.
The \kidsdata \ data is based on observations made with ESO Telescopes at the La Silla Paranal Observatory under programme IDs 177.A-3016, 177.A-3017, 177.A-3018 and 179.A-2004, and on data products produced by the KiDS consortium. The KiDS production team acknowledges support from: Deutsche Forschungsgemeinschaft, ERC, NOVA and NWO-M grants; Target; the University of Padova, and the University Federico II (Naples).
We used the gold sample of weak lensing and photometric redshift measurements from the fourth data release of the Kilo-Degree Survey (\cite{kuijken_fourth_2019}, \cite{wright_photometric_2020}, \cite{hildebrandt_kids-1000_2021}, \cite{giblin_kids-1000_2021}), hereafter referred to as \kidsdata. Cosmological parameter constraints from KiDS-1000 have been presented in \cite{asgari_kids-1000_2021} (cosmic shear), \cite{heymans_kids-1000_2021} (3×2pt) and \cite{troster_kids-1000_2021} (beyond $\Lambda$CDM), with the methodology presented in \cite{joachimi_kids-1000_2021}. Fabian Hervas Peters acknowledges support from the Centre National d\'Etudes Spatiales.
\end{acknowledgements}

%
%
\bibliography{first_ZOTERO,actual_paper}
\bibliographystyle{aa2}
\begin{appendix}

\section{Degeneracy with baryonic feedback parameters}\label{appendix:baryons}
\begin{figure}

    \includegraphics[width=3.4in]{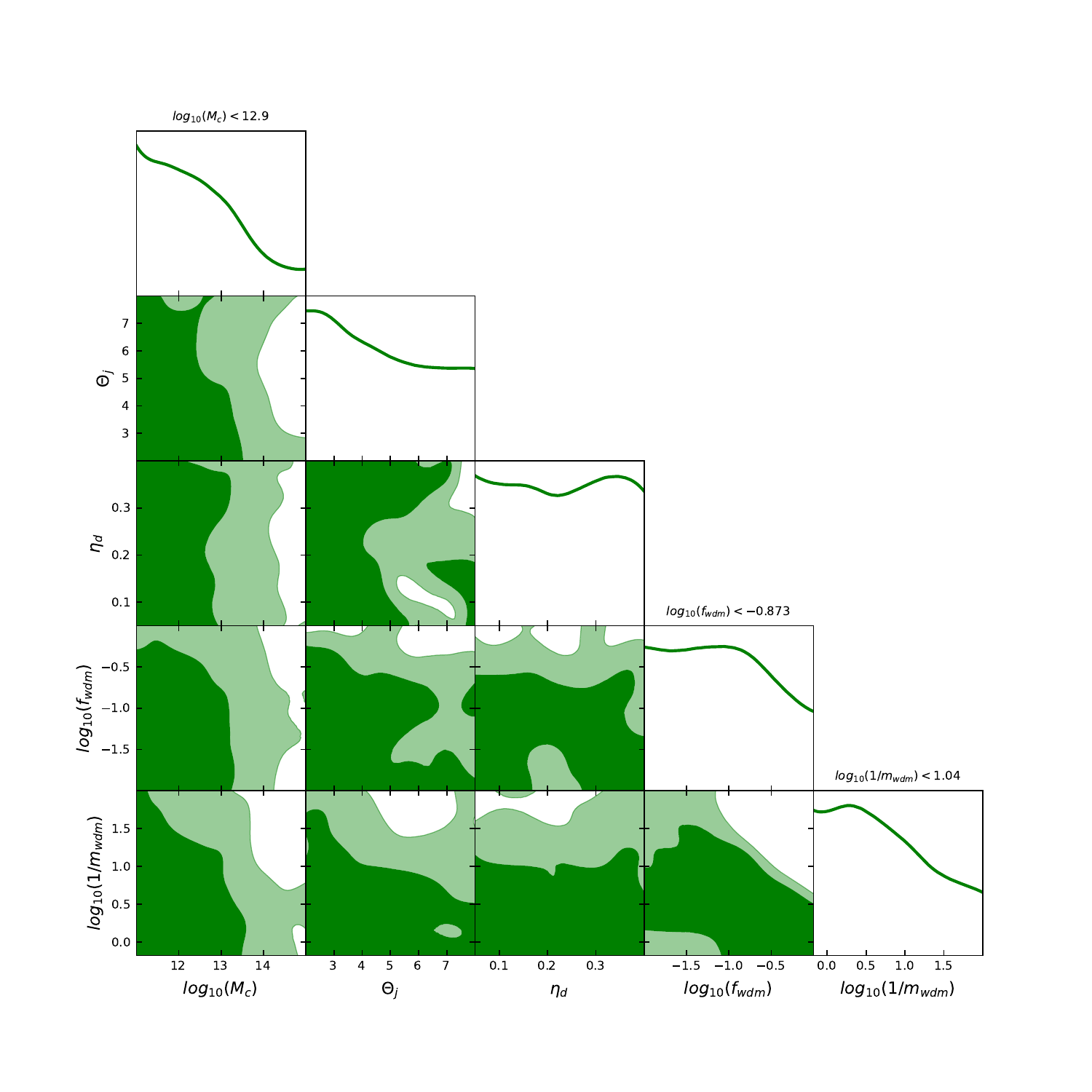} %
    \caption{Posterior contours of the baryonic feedback and MDM Parameters at the 68 and 95 \% confidence level. The baryonic feedback parameters are prior-dominated.
}

\label{fig:BF_MDM}
\end{figure}

It was shown in \cite{parimbelli_mixed_2021} that the baryonification scheme can be considered independently from the MDM parameters up to k=5h/Mpc at the percent level precision. As described in Section~\ref{sec:CSsig} the baryonic feedback parameters were treated as separate from MDM effects. Effectively, broad priors on our baryonic feedback parameters allowed for a large range of feedback suppressions. This broad prior was necessary to provide conservative constraints on the MDM parameters. We show in Fig.~\ref{fig:BF_MDM} that the baryonic feedback parameters are mainly prior dominated. Recent works use external datasets to constraint baryonic feedback suppresions, such as X-ray and kSZ data \citep{schneider_constraining_2022}, tSZ data \citep{pandey_inferring_2023} or gas density profiles from deep X-ray observations \citep{grandis2023determining}. This kind of study should improve the constraints obtained on the MDM parameters, as it would allow to disentangle baryonic processes from free-streaming DM effects. We leave this investigation to future work. 

\end{appendix}
\end{document}